\documentclass[prb, twocolumn]{revtex4-1}
\usepackage{times}
\usepackage{amsmath}
\usepackage[dvipdfmx]{graphicx}
\graphicspath{{/Users/Yo/Library/Mobile\ Documents/rsc/real_spin_texture/}, {/Users/Yo/Pictures/}}
\usepackage{subfigure}
\usepackage{here}
\usepackage{url}
\begin{document}
\title{Large Anomalous Nernst Effect in a Skyrmion Crystal}
\date{\today}
\author{Yo Pierre Mizuta}
\email{mizuta@cphys.s.kanazawa-u.ac.jp}
\affiliation{Graduate School of Natural Science and Technology, Kanazawa University, Kakuma, Kanazawa, 920-1192 JAPAN} %\\
\author{Fumiyuki Ishii}
\affiliation{Faculty of Mathematics and Physics, Kanazawa University, Kakuma, Kanazawa, 920-1192 JAPAN} %\\
\begin{abstract}
%Within the semiclassical Boltzmann transport theory, the formula 
%for Seebeck coefficient $S$ is derived for an isotropic two-dimensional
Thermoelectric properties of a model skyrmion crystal were theoretically investigated, and it was found that its large anomalous Hall conductivity, corresponding to large Chern numbers induced by its peculiar spin structure leads to a large transverse thermoelectric voltage through the anomalous Nernst effect. This implies the possibility of finding good thermoelectric materials among skyrmion systems, and thus motivates our quests for them by means of the first-principles calculations as were employed here.
\end{abstract}

\keywords{thermoelectric effect, skyrmion, anomalous Hall effect, anomalous Nernst
effect, Berry curvature, two-dimensional electron gas}

\maketitle

\section{Introduction}
Thermoelectric (TE) generation, harvesting waste heat and turning it into electricity, should play an important role in realizing more energy-efficient society and overcoming the global warming. Nevertheless, it is yet to be widely used, mainly due to its still limited efficiency. 
There have been many studies pursueing highly efficient TE systems with a large value of figure of merit: $Z_X=\sigma X^2 /\kappa$, where $\sigma$ and $\kappa$ are longitudinal electrical and thermal conductivity respectively, and $X=S\ {\rm or}\ N$ is the Seebeck or Nernst coefficient depending on whether we use the longitudinal or transverse voltage for power generation. Here we omitted labels $xx$ or $yy$ in $\sigma$ and $\kappa$ by assuming an isotropic system.
Among those, our study sheds light on the anomalous effect of electrical conduction perpendicular to an electric field (anomalous Hall effect, AHE) \cite{nagaosaanomalous2010} or to a temperature gradient (anomalous Nernst effect, ANE) \cite{xiaoberry-phase2006} on TE performance, focusing on a particular contribution to the conductivity of AHE (ANE), namely the so-called \textit{intrinsic} term $\sigma^{\rm int}_{xy}(\alpha^{\rm int}_{xy})$ expressed as a functional of Berry curvature\cite{xiaoberry2010} $\Omega({\bf k})\equiv i \langle \partial_{\bf k}u |\times |\partial_{\bf k}u \rangle$ in momentum({\bf k}) space (See the middle line of Eq.(\ref{sigalp}) in Appendix A), where $|u_{\bf k} \rangle$ is the periodic part of a Bloch state.
%$ in momentum space intrisic to perfect crystalline electronic structures. Since $\Omega){\bf k})$ is a key quantity in topological insulators(TIs), we previouly discussed the prototypical case of 2D electron gas with Rashba spin-orbit interaction ~cite{} or with Dirac-type dispersion, both of which are expected to appear in magnetic interfaces involving TIs.

%  Among those works were the
%prediction of advantage of lower dimensional systems, 
Systems hosting AHE/ANE are found in magnetic materials, both normal semiconductors \cite{culceranomalous2003} and topological insulators \cite{changexperimental2013}. 
%%%%%%%%%%%% Our prev. works%%%%%%%%%%%%%
We previously studied simple models of 2D electron gas in an interface composed of those materials \cite{Our-RZ, Our-DZ}.
The anomalous effect on Seebeck coefficient was fairly large there, but remained rather small compared to what will be reported in this paper, which can mainly be attributed to the limited magnitude of the anomalous Hall conductivity (AHC) $\sigma^{\rm int}_{xy} \le 1$ (The unit is taken to be $e^2/h$ for the AHC in 2D in this paper) there.  
%%%%%%%%%%%% Our prev. works%%%%%%%%%%%%%
%There, however, the conductivity of AHE(AHC) usually remains in the order of one unless we consider multilayer systems  or unrealistically long-range hopping processes[].

There have been several ideas proposed for obtaining large AHC, such as the manipulation of massive Dirac cones (each of them being the source of $\sigma^{\rm int}_{xy}=1$) by controlling parameters \cite{fanglarge2014}, or the extention of system dimension in the normal-to-plane direction (2D to 3D) \cite{jiangquantum2012}. 

What we focus here is another one, namely, the control of real-space spin textures %%rev. 160105%%%
which are well known to induce another contribution to the AHE/ANE, often called the {\it topological} Hall/Nernst conductivity\cite{neubauertopological2009} $\sigma^T_{xy}(\alpha^T_{xy})$ that also has a geometrical meaning related to $\sigma^{\rm int}_{xy}(\alpha^{\rm int}_{xy})$.  In terms of Berry-phase theory \cite{xiaoberry2010}, the emergence of {\it topological} terms in the continuum limit (spin variation scale $\gg$ atomic spacing), is an analogue of the ordinary Hall effect with the external $B$ field just replaced by \textit{spin magnetic field} $B_{\rm spin}$, which is the real-space Berry curvature $\Omega({\bf R})$ itself, proportional to a quantity called \textit{spin scalar chirality} $\chi_{ijk}\equiv {\bf m}_i \cdot {\bf m}_j \times {\bf m}_k$ reflecting the geometrical structure spanned by each spin trio $({\bf m}_i,\ {\bf m}_j,\ {\bf m}_k)$. 
Although it should be better to treat $\sigma^{\rm int}_{xy}$ and $\sigma^{T}_{xy}$ on an equal footing, we will omit the latter (\textit{topological}) term in the main part of this paper because the validity of the simple relation $B_{\rm spin} \propto \Omega({\bf R})$ is not clear for our system of rather short spin variation scale. Still, the omission can be justified within this approximation (See Appedix B for details). 

Among many possible spin textures, what we target here is the so-called skyrmion crystal (SkX) phase observed even near room-temperature \cite{yunear2010}, where skyrmions, particle-like spin whirls, align on a lattice.
The skyrmionic systems, originally explored in nuclear physics \cite{skyrm1962}, have been studied extensively these days in condensed matter physics as well. Typical SkX-hosting materials are some of the transition metal silicides/germanides: MnSi \cite{muhlbauer2009}, MnGe, FeGe, or heterostructures such as monolayer Fe on Ir(111) \cite{heinzespontaneous2011}, in all of which the Dzyaloshinskii-Moriya term, a spin-orbit coupling effect peculiar to their inversion-asymmetric crystal structure, plays a crucial role in the emergence of skyrmions. 

Regarding the AHE in skyrmionic systems, phases with a quantized AHC, i.e. $\sigma^{\rm int}_{xy}=C(e^2/h)$ with an integer $C$ called Chern number, have been recently predicted in the models of SkX where the conduction electron spins are exchange-coupled to the SkX either strongly \cite{hamamotoquantized2015} or weakly \cite{ladoquantum2015}, preceded by a report of QAHE in a meron (half-skyrmion) crystal\cite{mccormicktuning2015}.
In this paper we focus on the case picked out in Ref.[\onlinecite{hamamotoquantized2015}] because of its particularly large $\sigma^{\rm int}_{xy}$, implying the possibility of large ANE as well, thanks to the close relation between AHE and ANE \cite{xiaoberry-phase2006, miyasatocrossover2007}, and that is the main point we confirmed in this study. 

At the end of this section, we stress that the computational method used in our study is based on first-principles, which can be applied to the exploration of realistic materials of SkX etc. from the same perspective. 
\section{Expressions of thermoelectric quantities}
The formulae for the thermoelectric cofficients to be evaluated follow from the linear response relation of charge current: ${\bf j}=\tilde{\sigma}{\bf E}+\tilde{\alpha}(-\nabla T)$, where {\bf E} and $\nabla T$ are the electric field and temperature gradient present in the sample.  

Using the conductivity tensors $\tilde{\sigma}=[\sigma_{ij}]$ and $\tilde{\alpha}=[\alpha_{ij}]$, we obtain
\begin{equation} \left\{
 \begin{array}{l}
{\displaystyle S\equiv S_{xx} \equiv \frac{E_x}{\partial_xT}=\frac{S_0 + \theta_H N_0}{1+\theta_H^2}}
 \vspace{2mm}\ \  \ \ \ \\ %{\rm for}\ \ \ \  i=x \ \ {\rm or}\ \  y \\
{\displaystyle N \equiv S_{xy} \equiv \frac{E_x}{\partial_y
 T}=\frac{N_0 - \theta_H S_0}{1+ \theta_H^2}=-S_{yx}.} \label{SN}
\end{array} \right.
\end{equation}
Here we defined $S_0 \equiv \alpha_{xx}/\sigma_{xx}, \theta_H \equiv
\sigma_{xy}/\sigma_{xx},\ N_0 \equiv \alpha_{xy}/\sigma_{xx}$ for a simpler notation. See Eq.(\ref{sigalp}) in Appendix A for the specific form of the conductivity tensors $\sigma$ and $\alpha$ we consider here. Throughout this paper, since our 2D system has no periodicity in $z$ direction, we discuss $\sigma_{ij}\equiv \sigma_{ij}^{2D}\equiv d\times \sigma_{ij}^{3D}$, which is independent of the film thickness $d$ and has the dimension of $e^2/h$ whose unit can be $\Omega^{-1}$, instead of {\it real} conductivity $\sigma_{ij}^{3D}$.

Note that the Seebeck coefficient $S_0$ estimated without considering Berry
curvature is obtained by setting $\theta_H=0$ and $N_0=0$ in
Eq. (\ref{SN}).

\section{Model}
\begin{figure}
	\begin{center}
 		\includegraphics[width=0.35\textwidth]{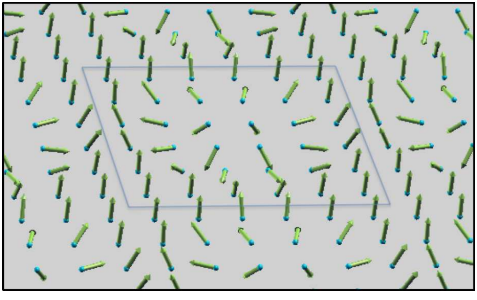} 
	\end{center}
	\caption{A view of an unit cell of $6\times 6$ SkX.}
	\label{skx}
\end{figure}

We consider a magnetic SkX on a two dimensional square lattice, with its unit cell of lattice constant $2\lambda=19.8 {\mathrm \AA}$ containing $6\times 6$ spins, thus the atomic lattice spacing $a=2\lambda/6$ being 3.3\AA. 
% \footnote{For $n \times n$ SkX, the atomic lattice spacing $a$ is given by $a=2\lambda/n$, the value of which affects the hopping amplitude among atomic sites.}.
 The spin configuration is equivalent to the one studied in Ref.[\onlinecite{hamamotoquantized2015}], i.e., the spherical coordinates of spin ${\bf m}({\bf r}_i)$ located at site $i$  are set as $\theta_i=\pi(1-r_i/\lambda)$ for $r_i<\lambda$ and $\theta_i=0$ for $r_i>\lambda$, along with $\phi_i=\tan^{-1}(y_i/x_i)+C$, where $C$ is an arbitrary constant.  In order to simulate the simplest case, we assumed each spin is that of a hydrogen atom. The spin modulation in the system is shown in Fig.\ref{skx}.

\section{Computational procedure}
Our calculations consist of three steps: (1) Obtain the electronic states of target SkX using {\it OpenMX} code \cite{openmx}, (2) Construct maximally localized Wannier functions (MLWF) employing {\it Wannier90} code\cite{wannier90}  and finally (3) Compute from the obtained MLWF all the necessary transport quantities [tensors $\sigma$ and $\alpha$ in Eq.(\ref{sigalp})] expressed according to the Boltzmann semiclassical transport theory, where we adopted constant-relaxation-time approximation with a fixed value of $\tau=0.1$ps \footnote{This value is expected to be realistic \cite{shurelectron1996}, but we will find later in the discussions that our results are very sensitive to the choice of $\tau$.}.

In step(1), one \textit{s}-character numerical pseudo-atomic orbital with cutoff radius of 7 bohr was assigned to each H atom. The present calculation for a non-collinear magnetic system was realized by applying a spin-constraining method \cite{kurzabinitio2004} in the non-collinear density functional theory \cite{kublerdensity1972}. Step(1) yielded 6$\times$6=36 non-spin-degenerate occupied bands and the equal number of unoccupied ones, among which only the former 36 bands were used in step(2) to construct MLWF and interpolated with them to calculate the conductivities. 
 In step(3), two modules were used in {\it Wannier90}: {\it berry} module based on the formalism in Ref.[\onlinecite{wangwannier2006}] to compute $\sigma^{\rm int}_{xy}$ and {\it boltzwann} module introduced in Ref.[\onlinecite{pizziboltzwann2014}] to compute $\sigma_{xx}$ and $S_0$, in both of which the sampling for integrations was performed on $50\times 50$ {\bf k}-points. Besides these, numerical intergrations were carried out to evaluate $\alpha_{xy}$ from $\sigma_{xy}$ via Eq.(\ref{sigalp}).

The above procedure was tested\footnote{For the $4\times 4$ SkX, the calculated band structure and the Fermi energy dependence of AHC were in overall agreement with the ones reported in Ref.[\onlinecite{hamamotoquantized2015}], confirming the reproducibility of the similar situation by different approaches, i.e., tight-binding model as in Ref.[\onlinecite{hamamotoquantized2015}] and our first-principles treatment.}, referring to Ref.[\onlinecite{hamamotoquantized2015}].

\section{Results and Discussions}
\subsection{Electronic structure and conductivities}

\begin{figure}
	\begin{center}
%	\subfigure[$4\times 4,\ a=11.2\AA $]{%
%		\includegraphics[width=0.2\textwidth]{../real_spin_textures/skyrmion/4_11p2_b+AHC.eps}}
%	\subfigure[$4\times 4,\ a=13.2\AA $]{%
%		\includegraphics[width=0.2\textwidth]{../real_spin_textures/skyrmion/4_13p2_b+AHC.eps}} \\
%	\subfigure[$6\times 6,\ a=19.8\AA $]{%
%		\includegraphics[width=0.5\textwidth]{../real_spin_textures/skrmH/6_19p8_b+sigxx+ahc_combined.eps}
		\includegraphics[width=0.5\textwidth]{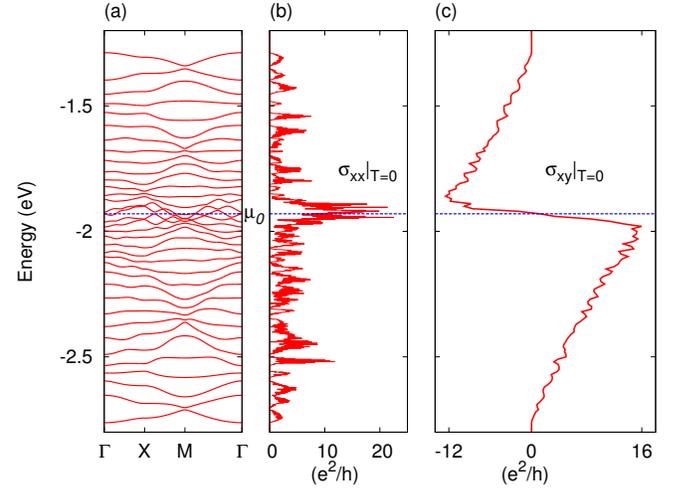}
	\end{center}
	\caption{(a) Band structure and Fermi-level dependence of (b) longitudinal and (c) anomalous Hall conductivity of $6\times 6$ SkX. The blue dashed line indicates the $\mu_0$ mentioned in the main text.}
	\label{b_ahc}

\end{figure}

First we show the obtained band structure of the 36 occupied states in Fig.\ref{b_ahc}(a).
%, together with the Fermi-energy dependence of AHC in each case.
%===== on BAND STRUCTURES ========
We notice there that
%4_4% , for all of the three cases, 
 each band is well isolated from (not touching with) each other, except for the four bands around the middle energy range [-2.0, -1.9]eV, which we shall hereafter refer to as ``central bands". We also see that some neighboring bands, including the center ones, tend to converge toward M point $(0.5,\ 0.5)\pi/(2\lambda)$. 
%4_4$ As we zoomed the dispersion, however, we found there was no exact degeneracy (touching) on the plotted lines for the $4\times 4$ cases. 
Away from M point, we see some degenerate points. Regarding the dispersions, quite nice symmetry with respect to the energy $\mu_0 \simeq -1.93$eV in the center bands also deserves close attention. Further analyses are needed in order to understand the secrets behind these interesting features.  

%===== on BAND STRUCtures ========
%===== on conductivities =======
Next, let us observe how the longitudinal conductivity $\sigma_{xx}$ and AHC at $T=0$K depend on the band filling (Fermi level) in Fig.\ref{b_ahc}(b) and (c) respectively.
The $\sigma_{xx}(\mu)|_{T=0}$ has a large peak in the central bands region, as can be expected from the obviously large density of states there. The $\sigma_{xx}$'s roughly symmetric variation with respect to $\mu_0$ (recognized when averaged over some spreaded energy) could have been anticipated intuitively from the above-mentioned symmetric band structure.% revealed in Fig.\ref{b_ahc}(a).    

The AHC, on the other hand, shows good anti-symmetric behavior with respect to $\mu_0$, which is quite understandable from the combined consideration of, again, the symmetric band structure and the assumption that each of the isolated (other than the central) bands is analogous to a Landau level formed by external magnetic field which contributes AHC=1 ($e^2/h$) in the quantum Hall effect.  The maximum absolute value of AHC is approximately 16, reached just below the central bands, while the second largest value of about 12 is located just above them. % $4\times 4$ and $6\times 6$, respectively.
%%% ? include or not ? %%%
%(Actually, this assumption is verified by evaluating band-resolved AHC (chern number of each band) using another computational method from the one employed to obtain the results in fig.\ref{b+AHC}, which in addition showed large (from 2 up to 7) chern numbers associated with four B0 bands for the $4\ times 4$ case.) 
%%% ? include or not ? %%%
These behaviors result in the drastical change of AHC around central bands as is prominent in Fig.\ref{b_ahc}(c). This very character has a strong effect on the TE properties of the system, as will be seen below.
%===== on conductivities========

%===== on TE properties ========
\subsection{Filling-dependence of thermoelectric properties}
\begin{figure}
	\begin{center}
		\includegraphics[width=0.5\textwidth]{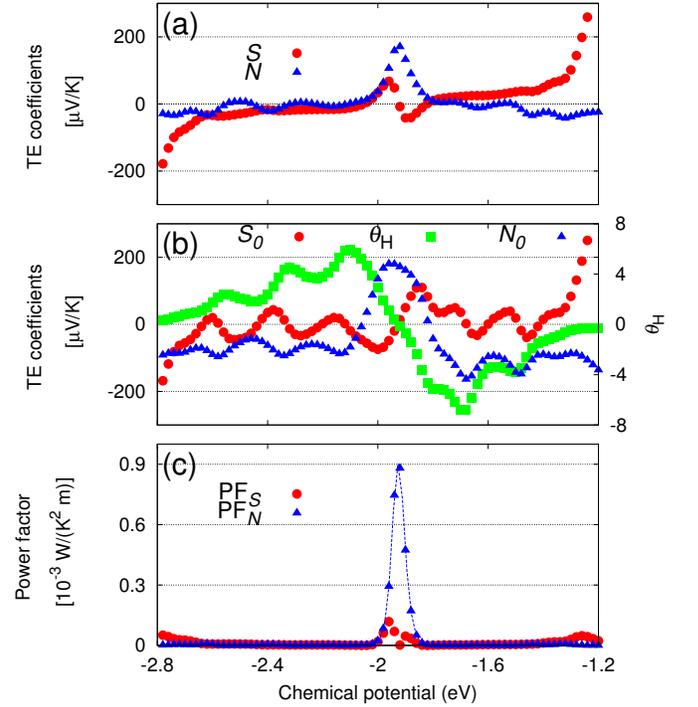}
	\end{center}
	\caption{Chemical potential dependence of the thermoelectric quantities of $6\times 6$ SkX at $T=300$K: (a)$S$ and $N$, (b)$S_0,\ N_0$(left axis) and $\theta_H$(right axis) [see Eq.(\ref{SN})], (c)Power factors corresponding to $S$ and $N$. The blue dotted line for PF$_N$ in (c) is drawn to guide the eye.} 
	\label{TE}
\end{figure}

We will proceed to our main subjects, the TE quantities of the system, especially focusing on their electron filling dependence, which we assume to be parametrized by the chemical potential $\mu$ within the rigid band approximation.
In all what is reported in this paper, the room temperature $T=300$K is assumed\footnote{Our choice $(a,\ T)=(3.3 \mathrm{\AA},\ 300{\rm K})$ is just an example of suitable sets for finding large $\alpha_{xy}$ in the following sense: If we obtain larger bandwidths, e.g. by compressing the lattice (making $a$ smaller), while somehow keeping maximally the shape of Berry curvature, $\sigma_{xy}'$ will be smaller, but thanks to the approximate relation $\alpha_{xy} \propto \sigma_{xy}'T$ [see Eq.(\ref{Mott})], at a higher temperature we may well get $\alpha_{xy}$ of the same magnitude as in the original system.}.
%The simplest ideal situation is when we have all the Hamiltonian matrix elements just multiplied by some constant $c$, i.e., $H^{(c)}=cH$, where $H^{(c)}$ and $H$ represent the modified and original Hamiltonian respectively, which will result in the relation $\sigma^{(c)}_{xy}(c\mu)=\sigma_{xy}(\mu)$, and if we write $\mu'=c\mu$, we obtain $d\sigma^c_{xy}(\mu')/d\mu'=(1/c)\sigma_{xy}(\mu)$ and finally $\alpha^{(c)}_{xy}(c\mu, cT)=\alpha_{xy}(\mu, T)$. \newline Intuitively, situations close to this could be achieved by adjusting the lattice spacing}
The Seebeck $S$ and Nernst $N$ coefficients, the terms $S_0, \ \theta_H,\ N_0$ constituting them [Eq.(\ref{SN})], and the power factors associated with each of $S$ and $N$, are shown in Fig.\ref{TE}(a), (b) and (c), respectively.  

%4_4% For the shorter lattice constant case in $4\times 4$ SkX (Fig.\ref{TE}(a)), $S$ is quite deviating from $\alpha$ in some part of the energy range,where $N$ attains quite a large value up to ~100 $\mu$V/K. 
%When the lattice is stretched to $a=13.2\AA$ (Fig.\ref{TE}(b)), a terrace of $N$ appears around the center of B0win. Let us note regarding the longitudinal counterpart $S$ that, there is an energy inside this range at which we have $S=\alpha$ (disappearance of anomalous effect), and the anomalous effect on $S$ tends to be strong around both ends of B0win or beyond. 
%These behaviors on the larger $a$ case become more pronounced for a larger $6\times 6$ SkX (Fig.\ref{TE}(c)), and therefore interpret them by referring to the more basic quantities shown in the same figure.  
As to the longitudinal $S(\mu)$, it is largely affected by the anomalous effect when $\theta_H(\mu)$ is finite (almost throughout the plotted $\mu$ range)%, except around the upper and lower edge, where $\theta_H \simeq 0$. 
We recognize in Fig.\ref{TE}(a,b) the following three points (i)-(iii) including their explanation based on what we have seen in Fig.\ref{b_ahc}: (i) $S({\mu_0})\simeq 0$ and $S_0({\mu_0})\simeq 0$, %with $\mu_0$ being located in the central bands. 
resulting from the combination of the above-mentioned behavior of $\sigma_{xx}|_{T=0}$ and $\sigma_{xy}|_{T=0}$, whose symmetrical centers came across with each other around $\mu_0$.  
(ii)  Just below and above $\mu_0$, $S$ shows the peaks (the higher one reaching $\approx 80\mu$V/K) of the sign opposite to that of $S_0$ despite rather small Hall angle $\theta_H$ there. This is thanks to the large $N_0$ that satisfies $|\theta_H N_0|>|S_0|$.  
(iii) Between the central and edge energy range, $S$ is strongly suppressed compared to $S_0$ due to large $\theta_H$. 
%While the latter is due to the symmetric $\sigma_{xx}|_{T=0}$ around the central bands, the former is  which is quite well mimicking the above-mentioned character of $\sigma_{xy}(\mu)$, is finite. Conversely, we find that $S \simeq S_0$ around both edges of the energy range, toward which $\theta_H$ diminishes. In addition, $S(\mu_0)=S_0(\mu_0)$ for the $\mu=\mu_0$, at which $\theta_H(\mu_0)=0$. It is also noticeable that 
%\propto \tau v_F^2D_F$, where $v_F$ and $D_F$ are the Fermi velocity and density of states at the Fermi level, respectively. This can be understood from the symmetric $D_F$ shown in Fig.\ref{b_ahc}(b).
%reflecting the  realized by the symmetric dispersion mentioned before). 
%4_4 All of these are qualitatively similar to the case of $4\times 4$. 
%Nevertheless, it is surprising that $|S|$ and $|S_0|$ attain (local) maximum values of \textit{opposite} signs ($|S|\simeq 80 \mu$V/K) just below and above $\mu_0$ despite rather small $\theta_H$ there, which is the result of strong anomalous effect $|S_0|<|\theta_H N_0|$ with large $N_0$.
%4_4 here than in $4\times 4$ case. 

On the transverse part $N$, what we notice in Fig.\ref{TE}(a,b) and their interpretations are the following two points (i) and (ii): (i) Around $\mu_0$, it shows a large peak $N \simeq N_0 \simeq 180 \mu$V/K. This is because of $\theta_H(\mu_0)\simeq 0$ [See Eq.(\ref{SN})] and large $\sigma_{xy}'(\mu_0)$ (the prime indicating the derivative with respect to $\mu$ at $T=0$) as a result of the large anti-symmetry of $\sigma_{xy}$ around $\mu_0$, which affects $N_0$ approximately through the Mott's relation [Eq.(\ref{Mott}), later to be discussed quantitavely] $N_0 \propto \sigma_{xy}'/\sigma_{xx}$.  (ii) Away from $\mu_0$, $N$ is strongly suppressed compared to $N_0$ due to large $\theta_H$, similarly to the relation between $S$ and $S_0$. This is different from the situation where $N_0$ is the far dominant contribution to $N$, as reported for example in Ref.[\onlinecite{hanasakianomalous2008}].  
%there is a sharp peak ~180$\mu$V/K around $\mu_0$, directly reflecting the shape of $N_0$, which in turn, is attributed to that of $\alpha_{xy}$.  
%Third, we comment on the compositions of $S$ and $N$. In the shorter-$a$ $4\times 4$ case, due to rather small $N$ but rather large $\theta_H$ especially in the energy range of [-1.5, -1.0]eV, $S \equiv (\alpha+\theta_H N_0)$ 
%4_4 Third, we comment on the compositions of $S$ and $N$. In the shorter-$a$ $4\times 4$ case, due to quite complicated ups and downs especially in $\alpha$, we cannot make any clear approximation in the expression of $S$ nor of $N$,
%$S \equiv (\alpha+\theta_H N_0)$ globally in the energy range of [-1.5,1.0]eV , although outside there $S\simeq \alpha$ thanks to suppressed values of both $\theta_H$ and $N_0$. Focusing on some of the largest values of $S$ and $N$ found above $\mu$~1.7eV, however, we can say that $S \simeq \alpha/(1+\theta_H^2)$ and $N\simeq -\alpha \theta_H/(a+\theta_H^2)$ hold.  For the largest $S$ and $N$ in long-$a$ case, on the other hand, $N\simeq N_0$, while $S\simeq \alpha/(1+\theta_H^2)$ is the same as for shorter-$a$.
%===== on TE properties ========
%\begin{figure}
%	\begin{center}
%		\includegraphics[width=0.4\textwidth]{../real_spin_textures/skyrmion/6_19p8_b+AHC.eps}
%	\end{center}
%	\label{bands6}
%	\caption{Band structures and Fermi-level dependence of anomalous Hall conductivity}
%\end{figure}
In view of the relation between $S$ and $N$, it is instructive that large $N$ appears thanks to the large asymmetry of $\sigma_{xy}$, at the same filling where $S$ diminishes due to the complete loss of asymmetry in $\sigma_{xx}$, showing that $\sigma_{xy}$ could be an additional freedom in seeking better thermoelectricity, when the material is properly designed.
%%% Confirmation of the Peak value of S and N by  Rough Evaluation of Mott's formula %%%
\subsection{Largest predicted thermoelectric voltages}
In order to better capture quantitatively the above-mentioned connection between conductivities and TE coefficients that leads to large values of the latter, which are seemingly attractive for TE applications, let us check whether the largest value $S_{\rm max}$ and $N_{\rm max}$ seen in the central energy range in Fig.\ref{TE}(a) can be roughly estimated via the well-known
Mott's formula, which says, 
\begin{equation}
	\alpha_{ij}=-\frac{\pi^2}{3}\frac{k_B^2T}{e}\sigma_{ij}^{'}(\mu)|_{T=0}, \label{Mott}
\end{equation}
where the prime in $\sigma_{xy}'$ means the derivative with regard to the Fermi energy (chemical potenatial $\mu$ at $T=0$).
%	The local maximum of $|S(\mu)|$ and $|N(\mu)|$, both in the range [-2.0, -1.9]eV, satisfy $\mu_S \lesssim \mu_N$ as found in Fig.\ref{TE}, and 
%we focus on the value $S(\mu_S)$ and $N(\mu_N)$ here. 
We write the chemical potential values that respectively give $S_{\rm max}$ and $N_{\rm max}$ as $\mu_S$ and $\mu_N$, both of which are close to $\mu_0$ but slightly different from each other.
Finding that $S_0(\mu_S)\approx 0,\ \theta_H(\mu_S)\approx 1$, $S$ reduces to  $S({\mu_S}) \simeq N_0(\mu_S)/2=\sigma_{xy}'/{2\sigma_{xx}}$. Similarly found is that $N(\mu_N) \simeq N_0(\mu_N)$ because of $S_0(\mu_N) \approx 0 \approx \theta_H(\mu_N)$.
A rough estimation of $\sigma_{xy}'(\mu_S)|_{T=0}$ can be obtained by linearly-approximating the drastical drop of AHC $\Delta \sigma_{xy}(\mu) \approx -28(e^2/h) \Omega^{-1}$ in the energy range $\Delta \mu \approx 0.13$ eV in Fig.\ref{b_ahc}(c). This gives $\sigma_{xy}'(\mu_S) \simeq \frac{\Delta \sigma_{xy}}{\Delta \mu} \approx 2.2 \times 10^{2}(e^2/h)$eV$^{-1}$. The other quantity we need is $\sigma_{xx}(\mu_S)$, which was found to be about $7.7(e^2/h)$. Combining these, a dimensionless factor of $x=k_BT/(\sigma_{xx}/\sigma_{xy}')\simeq 0.73$ is found at $T=300{\rm K}\ (kT\simeq 0.026{\rm eV}$). Finally our evaluation arrives at $N_{\rm max}=N(\mu_N)\simeq N(\mu_S) \simeq N_0(\mu_S) \simeq -\frac{\pi^2}{3}\frac{k_B}{e}x \approx 2 \times 10^2 \mu{\rm V/K}$ and therefore $S_{\rm max}=S(\mu_S) \approx 1 \times 10^2 \mu{\rm V/K}$.
These values are in fairly good accordance with the integration-derived values, hence clarifying that $T=300$K is \textit{low} enough (in comparison to the Fermi level) for the Mott's formula to be valid at least for rough estimation.

At this point we compare the maximum value of $\alpha_{xy} \approx 10^{-8}$ A/K corresponding to the above-estimated $N_{\rm max}$, with a value reasonably expected to be maximum within the low temperature approximation for the two-band Dirac-Zeeman (DZ) model we studied before \cite{Our-DZ}.
The low-$T$ approximated $\alpha_{xy}$ in Eq.(4) of Ref.[\onlinecite{Our-DZ}] can be rewritten as a 2D quantity $\alpha_{xy} \simeq (\pi k e/12h)(kT/\Delta)/\tilde{\mu}^2$, where $\Delta$ is the Zeeman gap and $\tilde{\mu}\equiv \mu/\Delta \ge 1$ for the electron-doped case.
Assuming rather small $\tilde{\mu}=2$, and choosing $(kT/\Delta)=0.3$ to loosely satisfy the low-$T$ criterion $kT \ll \mu-\Delta=\Delta$, we obtain $\alpha_{xy}\approx 10^{-10}$A/K for the DZ model, which is by two orders of magnitude smaller than in the $6 \times 6$ SkX.
Although some larger $\alpha_{xy}$ can be achieved beyond low-$T$ range, still larger values in the SkX should be ascribed to the AHC more than ten times larger than that of the DZ model.

Finally, since $N_{\rm max} \simeq 180\mu$V/K is a good enough value as a TE material, 
in order to further investigate the practical performance of the present system, we plot the powerfactor ${\rm PF}_S \equiv \sigma_{xx}S^2$ and ${\rm PF}_N \equiv \sigma_{xx}N^2$ in Fig.\ref{TE}(c). Note that, for the evaluation of power factors, unlike in the discussions of TE quantities up to now, we need to know $\sigma^{3D}_{xx}$. Therefore we assumed the SkX film thickness of 10nm. The maximum of ${\rm PF}_N \approx 10^{-3}$W/(K$^2$m) is comparable to the values of possible oxide TE candidates such as NaCo$_2$O$_4$ and ZnO \cite{ucsb_database}.
Although the thermal conductivity $\kappa$ was beyond the scope of this study, we just make a rough estimate here: Assuming the thermal conductivity $\kappa$ of the order of 1W/(Km) corresponding to good TE materials \cite{ucsb_database}, the present case realizes $Z_NT$ of the order of 0.3 at $T=300$K.

The main limitation of our results lies in the lack of knowledge on the processes through which electrons are scattered. 
Even if the single-$\tau$ approximation is valid, the magnitude of $\tau$ matters a lot. 
This is clearly manifested in the relation $N(\mu_N)\propto \tau^{-1}$. For example we obtain $N(\mu_N) \approx 20\mu$V/K if we suppose $\tau=1{\rm ps}$, although the Nernst coefficient of this size is still much larger than the so far reported values and is practically valuable \cite{sakuraba2016}.

%\subsection{Comment on the fixed parameters}

\section{Summary}
We computed the thermoelectric (TE) coefficients in a model of square skyrmion crystal, where spin-scalar-chirality-driven anomalous Hall/Nernst effects directly expressed in terms of Berry curvature in momentum space, were found to give rise to, not only a very different behavior in Seebeck coefficient from in the case of conventional TE materials, but also a surprisingly large Nernst coefficient. 
The future task is to understand the physics unique to such systems that leads to the prominent TE effect, and to find or design good TE materials among this class of materials using the first-principles method used here, which will also be of great practical importance for energy-saving applications. 

\begin{acknowledgments}
 The authors thank K. Hamamoto for providing us with detailed information on the model he and coworkers had studied in Ref.[\onlinecite{hamamotoquantized2015}].
This work was partially supported by Grants-in-Aid on Scientific Research under Grant Nos. 25790007, 25390008 and 15H01015 from Japan Society for the Promotion of Scienceand by the MEXT HPCI Strategic Program.
Computations in this research were performed using supercomputers at ISSP, University of Tokyo.
\end{acknowledgments}
 
% \section*{Acknowledgements}
%The authors thank the Yukawa Institute for Theoretical Physics at Kyoto University.
%Discussions during the YITP workshop YITP-W-13-01 on
%"Dirac electrons in solids" were useful to complete this work.
%Part of this research has been funded by the MEXT HPCI Strategic Program.
%This work was partly supported by Grants-in-Aid for Scientific Research
%(Nos. 25104714, 25790007, and 25390008) from the JSPS.

\appendix
\section{Expressions of conductivity tensors}
We derive below the semiclassical formulae for conductivity tensors.
The starting point is the expression for charge current ${\bf j}$
obtained in Ref.[\onlinecite{xiaoberry-phase2006}], which reads,  
\begin{align}
{\bf j}&=e\int \frac{d{\bf k}}{(2\pi)^2}g({\bf r},{\bf k}){\bf v}_{\bf k} \nonumber \\
 &+e\nabla_{\bf
 r}\times k_{\rm B}T({\bf r})\int \frac{d{\bf k}}{(2\pi)^2}{\bf \Omega}({\bf
 k})\log \left[1+\exp \left(-\frac{\varepsilon_{\bf k}-\mu}{k_{\rm B}T({\bf r})}\right) \right] \label{anomj},
\end{align}
where  $e,\ g({\bf r},{\bf k)},\ T({\bf r}),\ {\bf v}_{\bf k},\ \varepsilon_{\bf k},\ {\rm \Omega}({\bf k}),\ {\rm and}\  \mu$ stand for the electron's charge ($e<0$),  distribution function,  local temperature, (velocity, energy and {\bf k}-space Berry curvature) of an electron with wave number {\bf k} and chemical potential, respectively. 
%Hereafter we  use atomic units ($\hbar=1,\ |e|=1,\ m=1$) unless explicitly specified.

The second term is a correction that appears when spatial inhomogeneity [$T({\bf r})$ in the present case] exists. 
Simplifying the second term following Ref.[\onlinecite{xiaoberry-phase2006}] and substituting the set of equations of motion of a perturbed (by {\bf E} field)
Bloch electron
\begin{equation}
{\bf v}_{\bf k}=\frac{1}{\hbar}\frac{\partial \varepsilon_{\bf k}}{\partial {\bf k}}-\dot{\bf
 k}\times{\bf \Omega}({\bf k}),\ \ \dot{\bf k}=e{\bf E} \label{anomv}
\end{equation}
which was derived in Ref.[\onlinecite{sundaramwave-packet1999}], for ${\bf v}_{\bf k}$ and the form of distribution
\begin{equation}
g({\bf r},{\bf k})=\tau_{\bf k}{\bf v}_{\bf k}\cdot \left[e{\bf
 E}+(\varepsilon_{\bf k}-\mu)\left(-\frac{\nabla T}{T}\right)
 \left(-\frac{\partial f}{\partial \varepsilon}\right)\right],  \label{g}
\end{equation}
obtained as the solution of Boltzmann transport equation within relaxation
time ($\tau_{\bf k}$) approximation, for $g({\bf r},{\bf k})$, we obtain
\begin{equation}
\left\{\begin{array}{l}
{\displaystyle \sigma_{xx}=e^2\tau \int d{\bf k}\
 v_x({\bf k})^2\left(-\frac{\partial f}{\partial
			\varepsilon}\right),} \vspace{2mm} \\
{\displaystyle \sigma_{xy}=-\frac{e^2}{\hbar}\int d{\bf k}\
 \Omega_z({\bf k})=-\sigma_{yx},} \vspace{2mm} \\
{\displaystyle \alpha_{ij}=\frac{1}{e}\int d\varepsilon
\sigma_{ij}(\varepsilon)|_{T=0}\frac{\varepsilon-\mu}{T}\left(-\frac{\partial
						f}{\partial
						\varepsilon}\right) }  \ \ \ \   \\
\qquad \qquad{\rm for}\ \ \ \   i=x \ \ {\rm or}\ \  y,\  \ j=x\ \ {\rm or}\ \ y,%\vspace{2mm}\\
%%{\displaystyle \alpha_{ij}=\frac{1}{e}\int d\varepsilon
%\sigma_{ij}(\varepsilon)_{T=0}\frac{\varepsilon-\mu}{T}\left(-\frac{\partial
%						f}{\partial
%						\varepsilon}\right) } \
%%\ \  {\rm for}\  (i,j)=(x,y),\ (y,x)
\end{array}
\right. \label{sigalp}
\end{equation}
where  
$v_x({\bf k})\equiv \hbar^{-1}\partial \varepsilon({\bf k})/\partial k_x$ is the electron's group
velocity, and we assumed a constant relaxation time ($\tau_{\bf k}=\tau$).  
% in a band in which it resides
%We restricted ourselves to the case of isotropic 2DEG, i.e.,
%$\varepsilon({\bf k})=\varepsilon(k),\ {\bf \Omega}=(0,0,\Omega_z)$ and  
%We wrote  as  is , and $\epsilon_{ijk}$ is the completely antisymmetric tensor. 

%\section{Smaller Skyrmion}

%Use the \verb|\appendix| command if you need an appendix(es). The \verb|\section%| command should follow
% even though there is no title for the appendix (see above in the source of this %file).
\section{Comment on the topological Hall contribution}
We roughly estimate the magnitude of change in our result to be brought about by considering $\sigma_{xy}^T$ omitted in the main part of the paper.

The expression for $\sigma_{xy}^T$ in the Drude model treatment is,
\begin{equation}
	\sigma_{xy}^T=\frac{\omega_s\tau}{1+(\omega_s\tau)^2}\sigma_{xx},
\end{equation}
where $\omega_s \equiv eB_{\rm spin}/m_e$ with spin-scalar-chirality-induced field $B_{\rm spin}\equiv (\hbar/2e)\hat{\bf m}\cdot(\partial_x \hat{\bf m}\times \partial_y \hat{\bf m})$ determined by the unit vector of the background spin texture $\hat{\bf m}\equiv {\bf m}/|{\bf m}|$ and electron mass $m_e$.

Substituting the present SkX form of $\hat{\bf m}({\bf r})$, we obtain the spatially-inhomogeneous $B_{\rm spin}({\bf r})=(\hbar/2e)(\pi/r\lambda)\sin\pi(1-r/\lambda)$, which gives a skyrmion size $\lambda$-independent flux of $h/e$ \cite{hamamotoquantized2015}. Approximating by taking its unit-cell averaged value $\overline{B}_{\rm spin}=(h/e)/(\pi \lambda^2)$ that is roughly $10^3$ T in our case of $\lambda \approx 1{\rm nm}$, we obtain the frequency of $\overline{\omega_s}=(e\overline{B}_{\rm spin}/m_e \approx 2 \times 10^{14} {\rm s}^{-1}$ and thus $\overline{\omega_s}\tau \approx 20$ for the scattering-relaxation time of $\tau=0.1$ps assumed throughout this paper. Therefore, the contribution of $\sigma_{xy}^T \approx \sigma_{xx}/20$ adds less than 0.1 to $\theta_H$ and less than 0.1$S_0$ to $N_0$, which is good reason to omit $\sigma_{xy}^T$ in our case.

\bibliographystyle{apsrev4-1}
\bibliography{main_final}
\end{document}